\documentclass[fleqn,10pt]{SelfArx} 

\definecolor{color1}{RGB}{0,0,90} 
\definecolor{color2}{RGB}{0,20,20} 

\usepackage{hyperref} 
\hypersetup{hidelinks,colorlinks,breaklinks=true,urlcolor=color2,citecolor=color1,linkcolor=color1,bookmarksopen=false,pdftitle={Title},pdfauthor={Author},urlcolor=blue}

\usepackage{graphicx}
\usepackage[square]{natbib}
\usepackage{amsmath}
\usepackage{multirow}
\usepackage{subfigure}
\usepackage{pdflscape}

\newcommand{\n}[1]{\mathrm{#1}}

\JournalInfo{Published in IEEE Transactions on Magnetics, Vol. 47 (6), 1687-1692, 2011} 
\Archive{\href{http://dx.doi.org/10.1109/TMAG.2011.2114360}{DOI: 10.1109/TMAG.2011.2114360}} 

\PaperTitle{Improving magnet designs with high and low field regions} 

\Authors{R. Bj\o{}rk, C. R. H. Bahl, A. Smith and N. Pryds} 
\affiliation{\textit{Department of Energy Conversion and Storage, Technical University of Denmark - DTU, Frederiksborgvej 399, DK-4000 Roskilde, Denmark}} 
\affiliation{*\textbf{Corresponding author}: rabj@dtu.dk} 

\newcommand{\keywordname}{Keywords} 

\Abstract{A general scheme for increasing the difference in magnetic flux density between a high and a low magnetic field region by removing unnecessary magnet material is presented. This is important in, e.g., magnetic refrigeration where magnet arrays has to deliver high field regions in close proximity to low field regions. Also, a general way to replace magnet material with a high permeability soft magnetic material where appropriate is discussed. As an example these schemes are applied to a two dimensional concentric Halbach cylinder design resulting in a reduction of the amount of magnet material used by 42\% while increasing the difference in flux density between a high and a low field region by 45\%.}

\begin{document}

\flushbottom 

\maketitle 


\thispagestyle{empty} 

\section{Introduction}
Designing a permanent magnet structure that contains regions of both high and low magnetic field and ensuring a high flux difference between these can be challenging. Such magnets can be used for a number of purposes but here we will consider magnetic refrigeration as an example. Magnetic refrigeration is a potentially highly energy efficient and environmentally friendly cooling technology, based on the magnetocaloric effect. This effect manifests itself as a temperature change that so-called magnetocaloric materials exhibit when subjected to a changing magnetic field. In magnetic refrigeration a magnetocaloric material is moved in and out of a magnetic field, in order to generate cooling. The magnetic field is usually generated by permanent magnets \cite{Gschneidner_2008,Bjoerk_2010b}. In such magnet designs used in magnetic refrigeration it is very important to obtain a large difference in flux density between the high and the low flux density regions, between which the magnetocaloric material is moved in order to generate the magnetocaloric effect. This is because the magnetocaloric effect scales with the magnetic field to the power of 0.7 near the Curie temperature for most magnetocaloric materials of interest, and in particular for the benchmark magnetocaloric material Gd \cite{Pecharsky_2006,Bjoerk_2010d}. Because of this scaling it is very important that the magnetic field in a low field region is very close to zero. This is especially a problem in rotary magnetic refrigerators \cite{Bjoerk_2010b,Okamura_2007,Zimm_2007,Tusek_2009} where the high and low magnetic field regions are constrained to be close together. Here it is crucial to ensure that flux does not leak from the high field region into the low field region.

The permanent magnet structure can be designed from the ground up to accommodate this criterion, e.g., by designing the structure through Monte Carlo optimization \cite{Ouyang_2006}, or by optimizing the direction of magnetization of the individual magnets in the design \cite{Marble_2008,Choi_2008}. However, the resulting design may be unsuitable for construction. Here we present a scheme that applied to a given magnet design will lower the flux density in the low flux density region, thus increasing the difference in flux density, and lower the amount of magnet material used at the same time. No general way to improve the flux density difference for a magnet design has previously been presented.


\section{Physics of the scheme}\label{The optimization and improvement schemes}
The properties of field lines of the magnetic flux density can be exploited to minimize the magnetic flux in a given region. A field line is a curve whose tangent at every point is parallel to the vector field at that point. These lines can be constructed for any vector field. The magnitude of the magnetic flux density, $\mathbf{B}$, is proportional to the density of field lines. For a two dimensional problem, as will be considered here, with a static magnetic field, lines of constant magnetic vector potential, $A_\n{z}$, are identical to field lines of $\mathbf{B}$ if the Lorenz gauge, i.e. $\nabla{}\cdot{}\mathbf{A} = 0$, is chosen \cite{Haznadar_2000}. We begin by calculating a field line of the magnetic flux density, $\mathbf{B}$, i.e. an equipotential line of constant $A_\n{z}$, that encloses the area in which the flux density is to be minimized. All field lines enclosed by the calculated field line are confined to the enclosed area as field lines do not cross. These enclosed field lines are creating the flux density inside the calculated area. This procedure will only work for a two dimensional case, as in three dimensions a field line will not enclose a volume. Here a surface of field lines that enclose the volume must be used instead.

If we remove all magnet material enclosed within the chosen field line, it might be thought that no field lines should be present inside the area and the flux density should be zero. However, this is not precisely the case as by removing some magnet material the magnetostatic problem is no longer the same, and a new solution, with new field lines of $\mathbf{B}$, must be calculated. Thus a new field line that confines the area in which we wish to minimize the flux density can be found and the procedure can be iteratively repeated.

It must be made clear that the magnet material inside the calculated field line, i.e. the material that is removed, does contribute a non-zero flux density to areas outside the enclosing field line. This can be seen by considering each little piece of a magnet as a dipole, which will generate a flux density at any point in space. Thus by removing the enclosed magnet material the flux density will also be lowered in the high flux density region. However, this is more than compensated by the lowering of $||\mathbf{B}||$ in the low flux density region, due to the fact that the high flux density region is farther away from the removed material. This makes it possible to increase the \emph{difference} between the high and low flux density regions.

Field lines that do not pass through the high flux density region do not contribute to the flux density there. The scheme can also be used to remove the magnet material enclosed by these field lines.

The scheme must be run until a stopping criterion has been reached. This can be, e.g., that the flux density in the low flux density region has dropped below a certain value or that the volume of magnetic material has been reduced by a certain fraction. This is to ensure that the flux density in the high flux density region is not significantly reduced. In some cases successive applications of the scheme might result in removal of all magnet material. If, for example, one tried to remove the flux density on one side of an ordinary bar magnet by applying the scheme, one would simply remove slices of the bar magnet, until the magnet would be removed completely. This does result in zero flux density, but does not leave any region with flux at all.

As an additional improvement, the removed magnet material can be replaced by a high permeability soft magnetic material, to shield the low flux density area from field lines from the new magnet configuration. This will lower the flux density in the low flux density region further. If the magnet material is replaced by air the scheme is henceforth referred to as improvement scheme (Air), while if magnet material is replaced by soft magnetic material the reference term is improvement scheme (Iron). The difference between these two cases is illustrated in the next section.

Due to the high permeability of the soft magnetic material one would not necessarily have to replace all the enclosed magnet material with a soft magnetic material. Removing the magnet material and using only a small layer of soft magnetic material along the edge of the remaining magnet to shield the low flux density region will in general result in the same magnetic field as replacing all the magnet material with soft magnetic material. This will be an attractive option if the weight of the final assembly is an issue. However, the only difference between these two solutions is the amount of soft magnetic material used, and this option will not be considered further here.

\begin{figure}[!t]
\centering
\includegraphics[width=1\columnwidth]{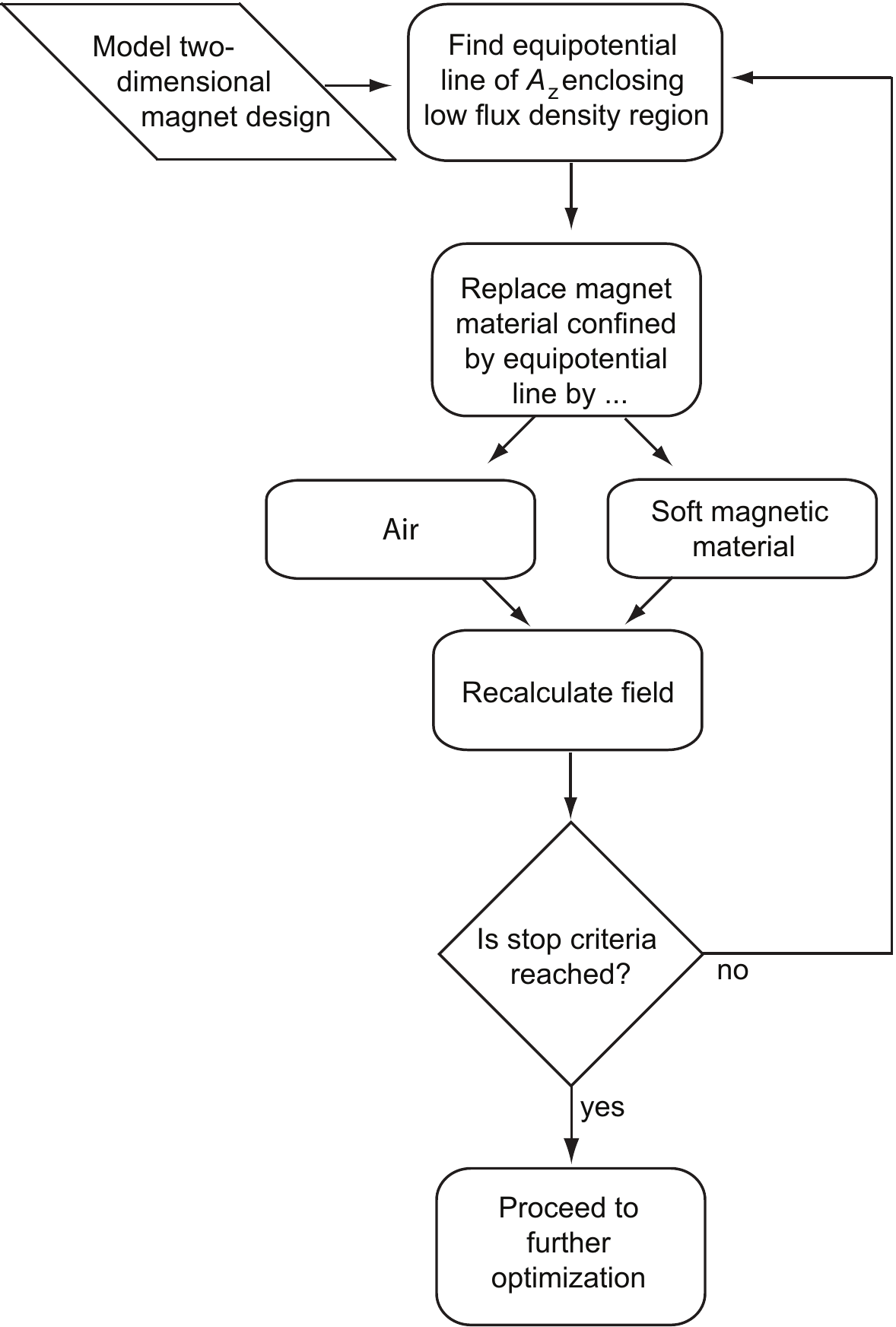}
\caption{The flow diagram for the improvement scheme.}\label{Fig.Schemes}
\end{figure}

In practice the scheme is implemented numerically and applied to a numerical simulation of a magnet design. The scheme is presented as a flow diagram in Fig. \ref{Fig.Schemes}.

\section{Applying the scheme}\label{The optimization and improvement schemes applied}
The improvement scheme is best illustrated through an example. Here we consider the concentric Halbach cylinder design, which is a cylindrical magnet with an air gap in between an outer and inner cylindrical magnet structure \cite{Bjoerk_2010a}. Each cylinder is magnetized such that the remanent flux density at any point varies continuously as, in polar coordinates,
\begin{eqnarray}
B_{\mathrm{rem},r}    &=& B_{\mathrm{rem}}\; \textrm{cos}(p\phi) \nonumber\\
B_{\mathrm{rem},\phi} &=& B_{\mathrm{rem}}\; \textrm{sin}(p\phi),\label{Eq.Halbach_magnetization}
\end{eqnarray}
where $B_{\mathrm{rem}}$ is the magnitude of the remanent flux density and $p$ is an integer \cite{Mallinson_1973,Halbach_1980}. The subscript $r$ denotes the radial component of the remanence and the subscript $\phi$ the tangential component. A positive value of $p$ produces a field that is directed into the cylinder bore, and a negative value produces a field that is directed outwards from the cylinder.

\begin{figure}[!t]
\centering
\subfigure[The full concentric Halbach cylinder.]{\includegraphics[width=1\columnwidth]{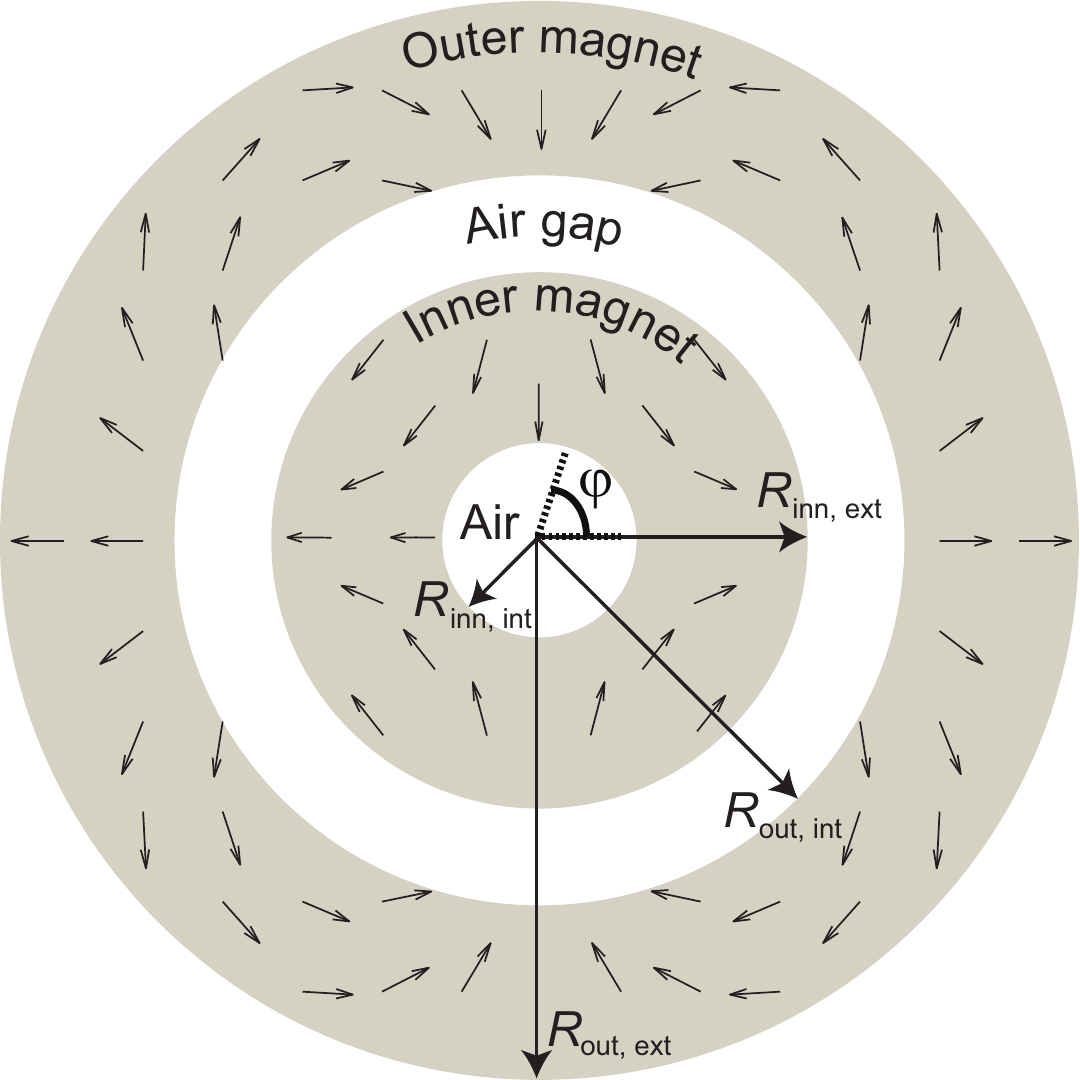}}\hspace{0.1cm}
\subfigure[A quadrant of the concentric Halbach cylinder design.]{\includegraphics[width=1\columnwidth]{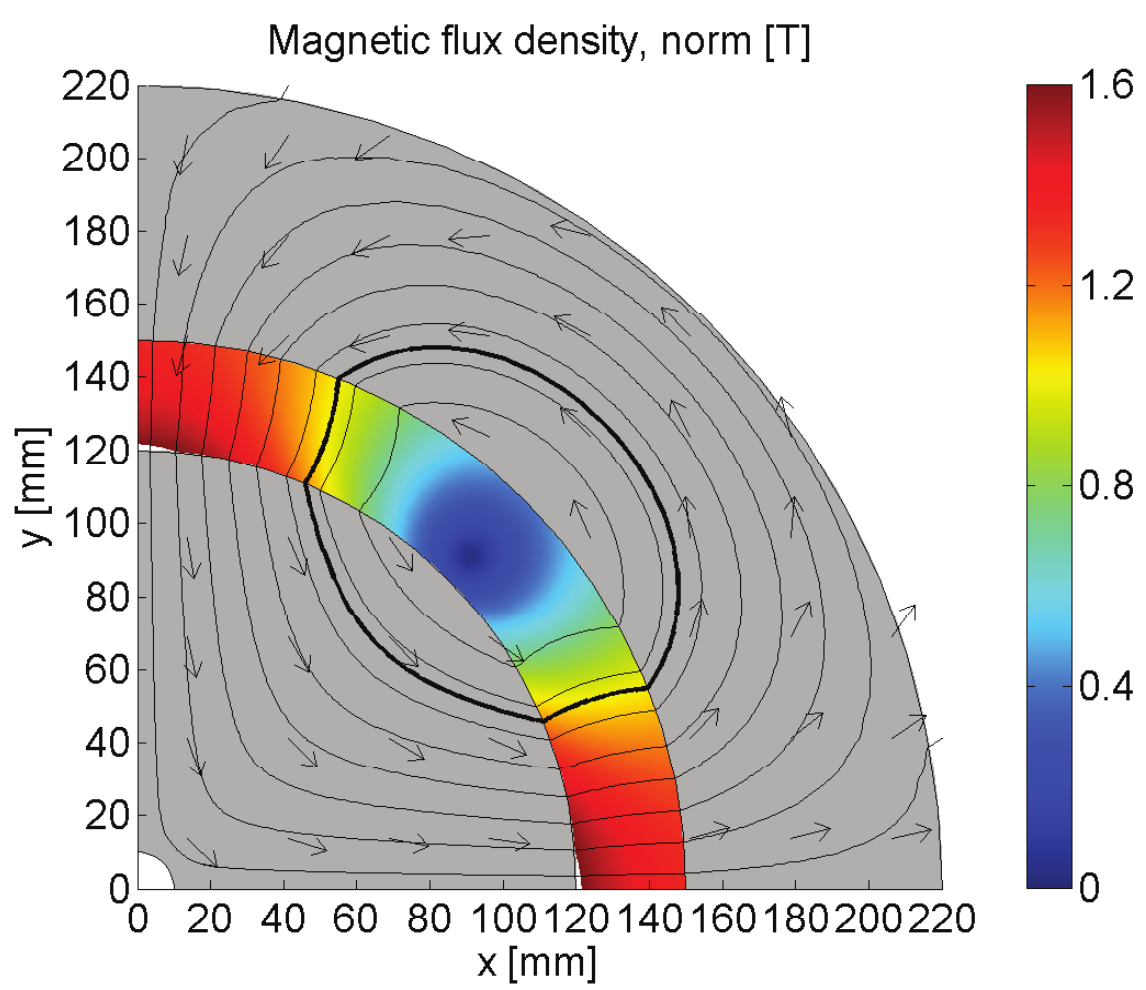}}\\
\caption{The full concentric Halbach cylinder (a) and a quadrant of the design (b). The magnetization is shown as black arrows on the magnets, which are grey. The flux density in the air gap between the cylinders is shown as a color map. In (b) the equipotential line of $A_\mathrm{z}$ which encloses the low flux density region is shown as a thick black line, whereas other contours of $A_\mathrm{z}$ are shown as thin black lines. It is magnet material inside the thick black line that is removed.}\label{Fig.Con_Hal_0-0-0-0-1-0-0-1_Air_Air}
\end{figure}

As an example we consider a magnet with four high and four low flux density regions which can be created by having a $p=2$ outer Halbach cylinder and a $p=-2$ inner Halbach cylinder and with dimensions $R_\mathrm{inn, int} = 10$ mm, $R_\mathrm{inn, ext} = 120$ mm, $R_\mathrm{out, int} = 150$ mm and $R_\mathrm{out, ext} = 220$ mm, which are indicated in Fig. \ref{Fig.Con_Hal_0-0-0-0-1-0-0-1_Air_Air}. The scheme could be equally well applied to any magnetic circuit with adjacent high and low flux density regions where the aim is to increase the difference between these regions.

In the example setup magnets with a remanence of $B_\mathrm{rem} = 1.4$ T and a relative permeability of $\mu_\mathrm{r} = 1.05$ which are the properties of standard neodymium-iron-boron (NdFeB) magnets \cite{Standard} are used. We define the high and low flux density regions to be of the same size and to span an angle of 45 degree each.

The improvement scheme will be applied to this design using a numerical two dimensional model implemented in the commercially available finite element multiphysics program \emph{Comsol Multiphysics} \cite{Comsol}.

As an equipotential line of $A_\mathrm{z}$ that encircles the low flux density region is chosen the equipotential line of $A_\mathrm{z}$ that goes through the point $r = 135 \;\mathrm{mm}, \phi{} = 22.5^{\circ}$, i.e. the point in the middle of the air gap, half way between the centers of the high and low flux density regions. This equipotential line is shown in Fig. \ref{Fig.Con_Hal_0-0-0-0-1-0-0-1_Air_Air}(b).

\begin{figure*}[!p]
\centering
\subfigure[Iteration 1.]{\includegraphics[width=1\columnwidth]{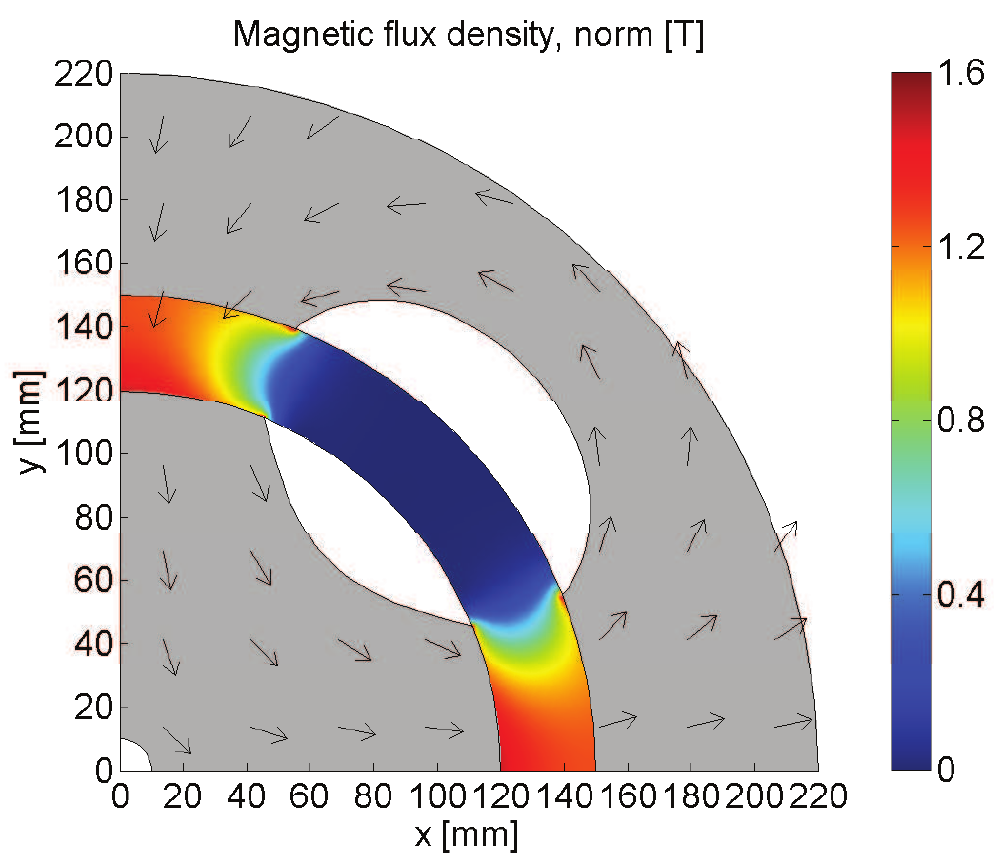}}\hspace{0.1cm}
\subfigure[Iteration 6.]{\includegraphics[width=1\columnwidth]{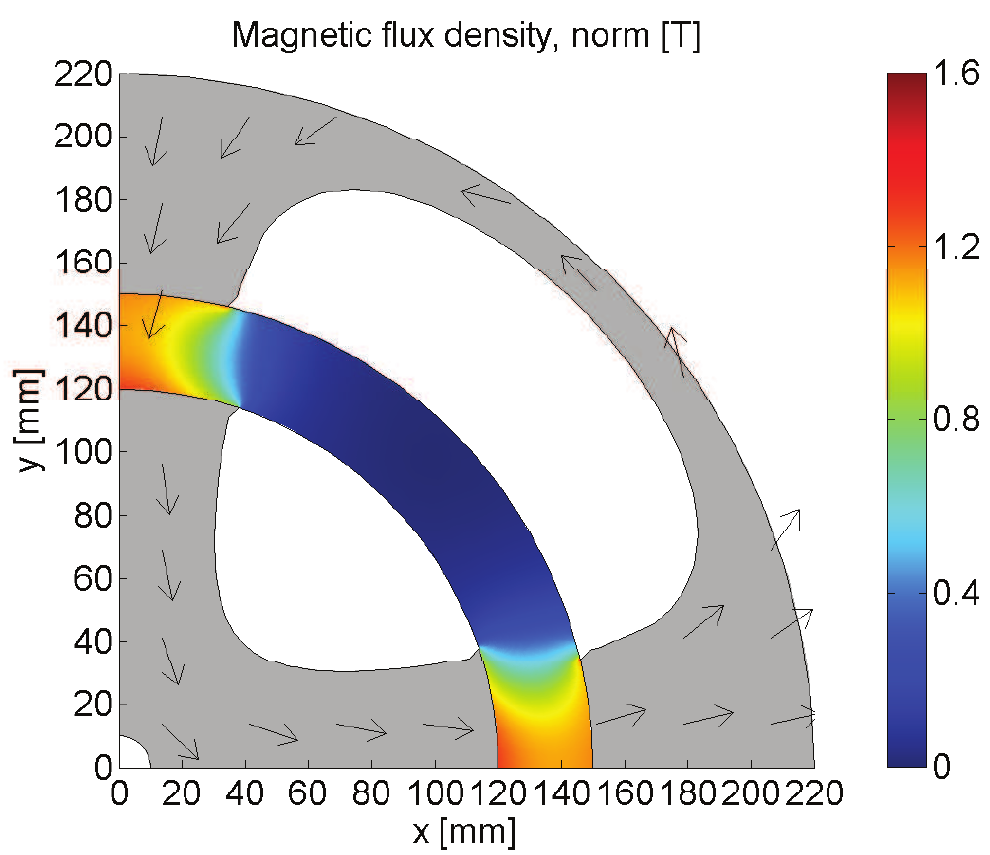}}\\
\caption{The improvement scheme (Air) applied to a quadrant of the magnet design. The first iteration step (a) and the sixth step (b) are shown. The first iteration step corresponds to Fig. \ref{Fig.Con_Hal_0-0-0-0-1-0-0-1_Air_Air}(b) where the magnet material enclosed by the thick black line has been removed.}\label{Fig.Optimization_scheme_pics_air}
\end{figure*}

\begin{figure*}[!p]
\centering
\subfigure[Iteration 1.]{\includegraphics[width=1\columnwidth]{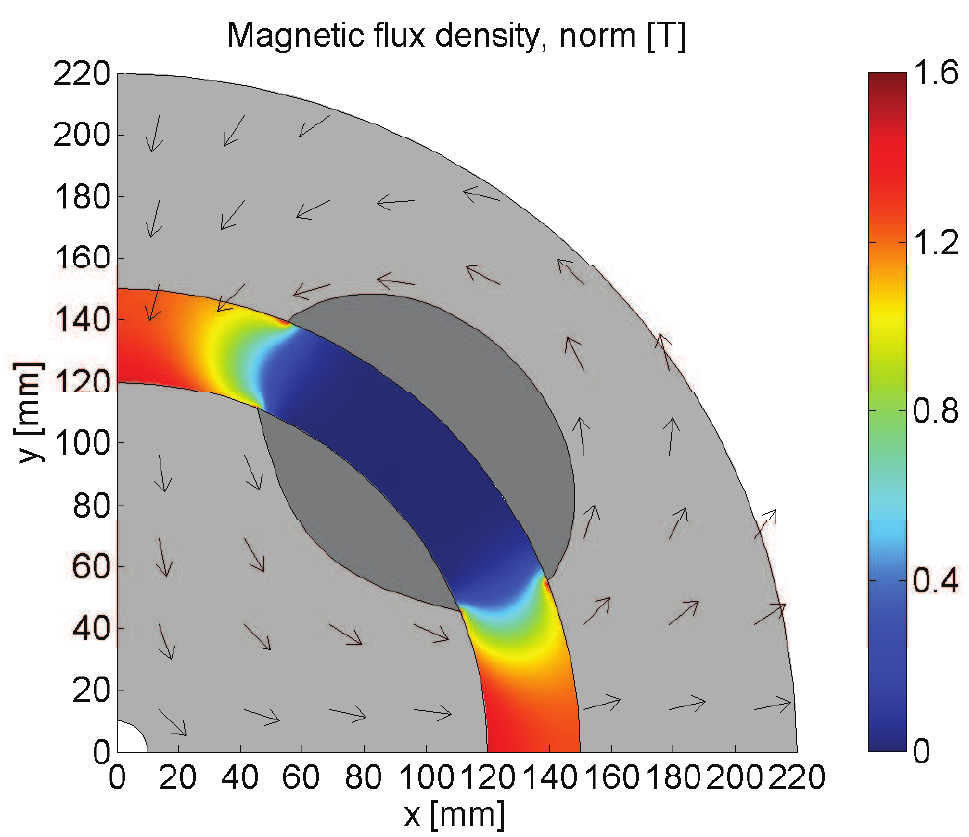}}\hspace{0.1cm}
\subfigure[Iteration 3.]{\includegraphics[width=1\columnwidth]{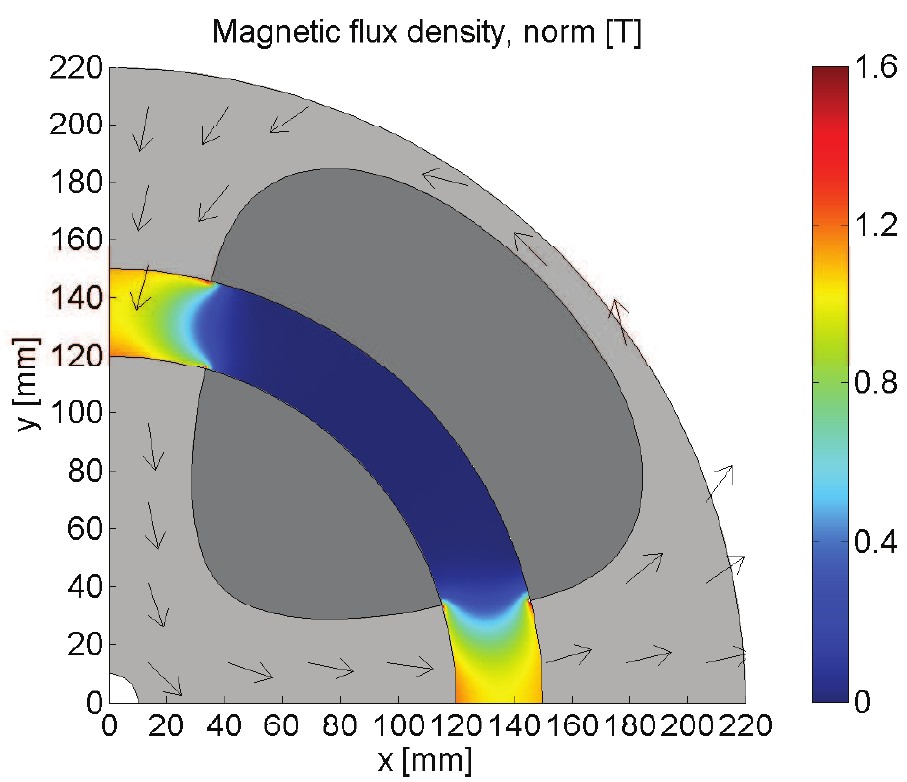}}\\
\caption{The improvement scheme (Iron) applied to a quadrant of the magnet design. The first iteration step (a) and the third step (b) are shown. The first iteration is identical to the first iteration in Fig. \ref{Fig.Optimization_scheme_pics_air}, expect that iron has been substituted instead of air. Areas of iron are indicated by dark grey.}\label{Fig.Optimization_scheme_pics_iron}
\end{figure*}

The improvement scheme in which the magnet material is replaced by air is shown in Fig. \ref{Fig.Optimization_scheme_pics_air}, while the same scheme where the magnet material is replaced by iron is shown in Fig. \ref{Fig.Optimization_scheme_pics_iron}. Iron was chosen as the soft magnetic material because it has a very high permeability as well as being easily workable and reasonably priced. 

It is seen that applying the improvement scheme does reduce the flux density in the low flux density region, but the flux density in the high flux density region also decreases as more and more magnet material is removed.

\begin{figure}[!t]
\centering
\includegraphics[width=1\columnwidth]{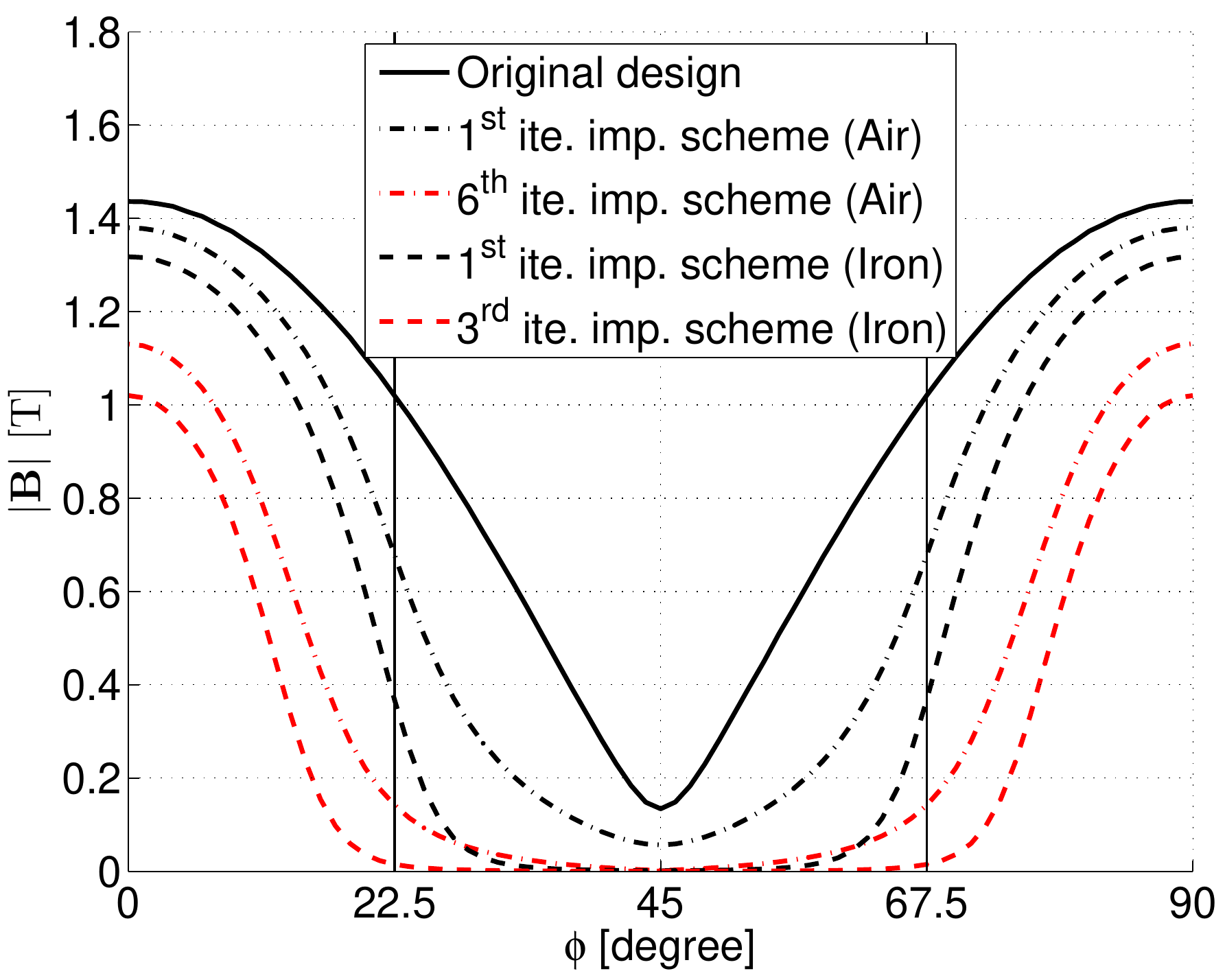}
\caption{The flux density as a function of angle in the middle of the air gap for the models shown in Figs. \ref{Fig.Optimization_scheme_pics_air} and \ref{Fig.Optimization_scheme_pics_iron}. The vertical lines separate the high and low flux density regions.}\label{Fig.Flux_function_of_angle_Converged_and_Improve}
\end{figure}

The effects of applying the two versions of the improvement scheme are shown in Fig. \ref{Fig.Flux_function_of_angle_Converged_and_Improve}, which shows the magnetic flux density in the middle of the air gap as a function of the angle, $\phi$.

It is seen from the figure that some flux is lost in the high flux density region, but the flux density in the low flux density region is also almost completely removed. Substituting with a soft magnetic material lowers the flux density in the low field region more than by substituting with air, because the soft magnetic materials shields the low field region.

\begin{figure}[!t]
\centering
\includegraphics[width=1\columnwidth]{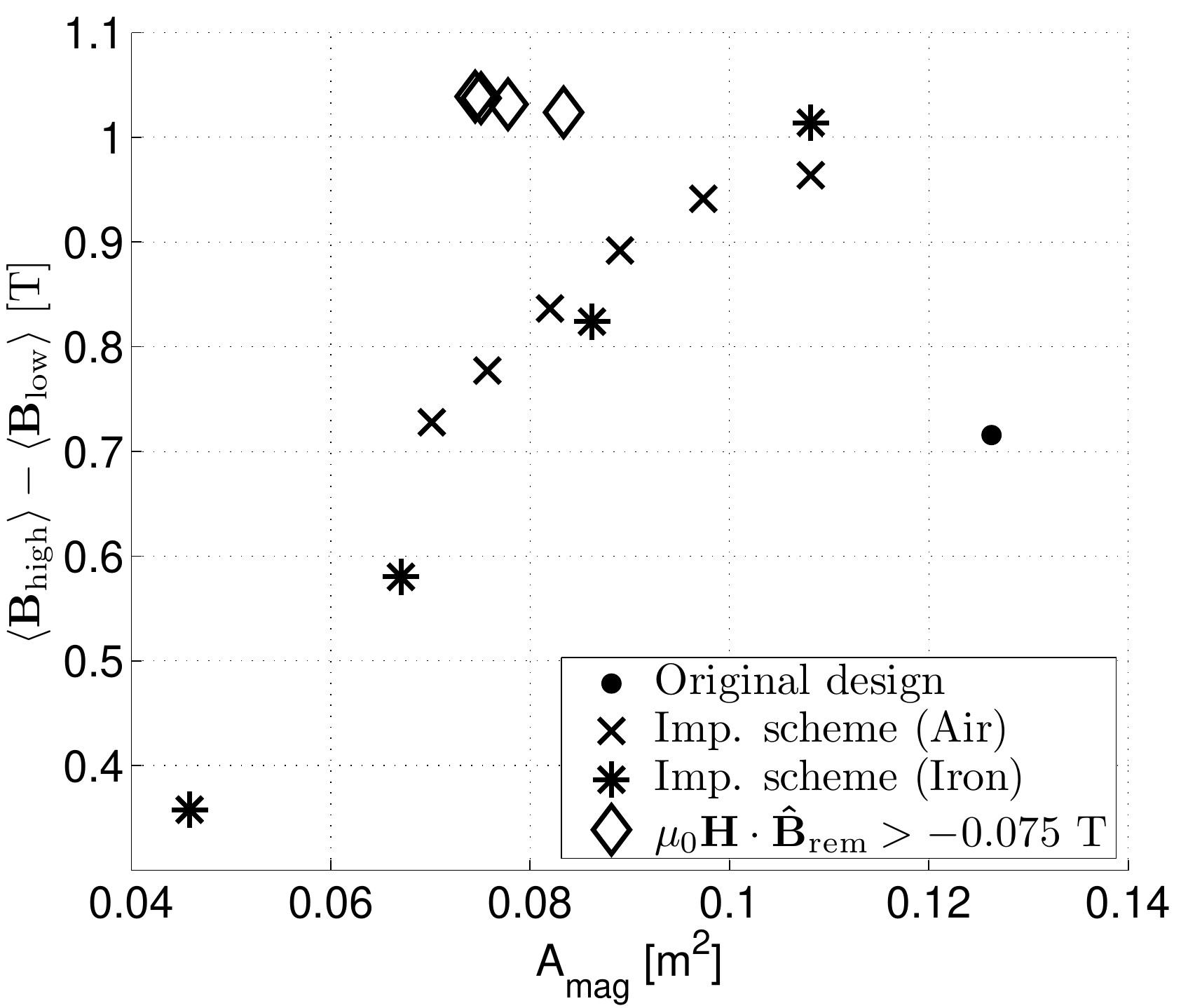}
\caption{The difference in flux density as a function of the cross-sectional area of the magnet for the improvement scheme. Decreasing values of $A_\mathrm{mag}$ indicates further iteration steps. Also shown is the difference in flux gained by replacing magnet material with a high permeability soft magnetic material where the applied magnetic field is parallel or almost parallel to the magnetization, i.e. where $\mu_0 \mathbf{H}\cdot{}\mathbf{\hat{B}_\n{rem}} > -\gamma$. This has been done on the model with a single application of the improvement scheme (Iron), i.e. the model with the highest flux difference.}\label{Fig.Iterations_improvements}
\end{figure}

The effect of applying the scheme is shown in Fig. \ref{Fig.Iterations_improvements} where the difference in flux density as a function of the cross-sectional area of the magnet is plotted. Both improvement schemes (Air) and (Iron) are shown. The data shown in Fig. \ref{Fig.Flux_function_of_angle_Converged_and_Improve}, but integrated over the complete high and low field regions are thus shown in this figure. As can clearly be seen applying the optimization schemes at first reduces the cross-sectional area of the magnet, $A_\mathrm{mag}$, while at the same time improving the difference in flux density between the high and the low flux density regions. The largest difference in flux density is obtained after only one iteration for both the improvement scheme (Air) and improvement scheme (Iron). In the latter case, which is also the best case, the amount of magnet material used is reduced by 15\% and the difference in flux density is increased by 41\%.

\clearpage

\section{Further design considerations}
The magnetic design produced by applying the improvement scheme might not be easily manufacturable, as is the case for the example considered above. Also, for this example the direction of magnetization varies continuously which is also unsuitable for manufacturing purposes. To overcome these problems the design must be segmented into regular pieces of permanent magnets, each with a constant direction of magnetization, and pieces of high permeability soft magnetic material. This segmentation can be accomplished in numerous ways, and is in itself a process that must be optimized. It must also be considered whether the added manufacturing cost of the magnet design is worth the increased difference in flux density and the lowered material cost. However, before segmenting a design an additional way of lowering the amount of permanent magnet material used in a given magnet design should also be considered.

As also stated in Ref. \cite{Bloch_1998} and \cite{Coey_2003} it is advantageous to replace magnet material with a high permeability soft magnetic material if the applied magnetic field is parallel to the remanence. In an ideal hard magnet the anisotropy field is infinite which mean that components of the magnetic field, $\mathbf{H}$, and $\mathbf{B}$ that are perpendicular to the direction of the remanence, $\mathbf{\hat{B}_\n{rem}}$, have no effect on the magnet. Here $\mathbf{\hat{B}_\n{rem}}= \mathbf{B_\n{rem}}/||\mathbf{B_\n{rem}}||$, i.e. the unit vector in the direction of $\mathbf{B_\n{rem}}$. Here we also propose to replace magnet material that has a small negative component of $||\mathbf{H}\cdot{}\mathbf{\hat{B}_\n{rem}}||$, as this has a poor working point far from the maximum energy density of the magnet. This will of course affect the flux density generated in the air gap, so care must be taken not to remove to much magnetic material. We thus propose to replace magnet material where
\begin{eqnarray}\label{eq.HparallelBrNew}
\mu_0 \mathbf{H}\cdot{}\mathbf{\hat{B}_\n{rem}} > -\gamma ~,
\end{eqnarray}
where $\gamma$ is a positive number. The value for $\gamma$ can be changed depending on the demagnetization curve for the magnet material being used, however in general $\gamma$ must be chosen small, i.e. on the order of at most 0.1 T.

\begin{figure}[!t]
\centering
\includegraphics[width=1\columnwidth]{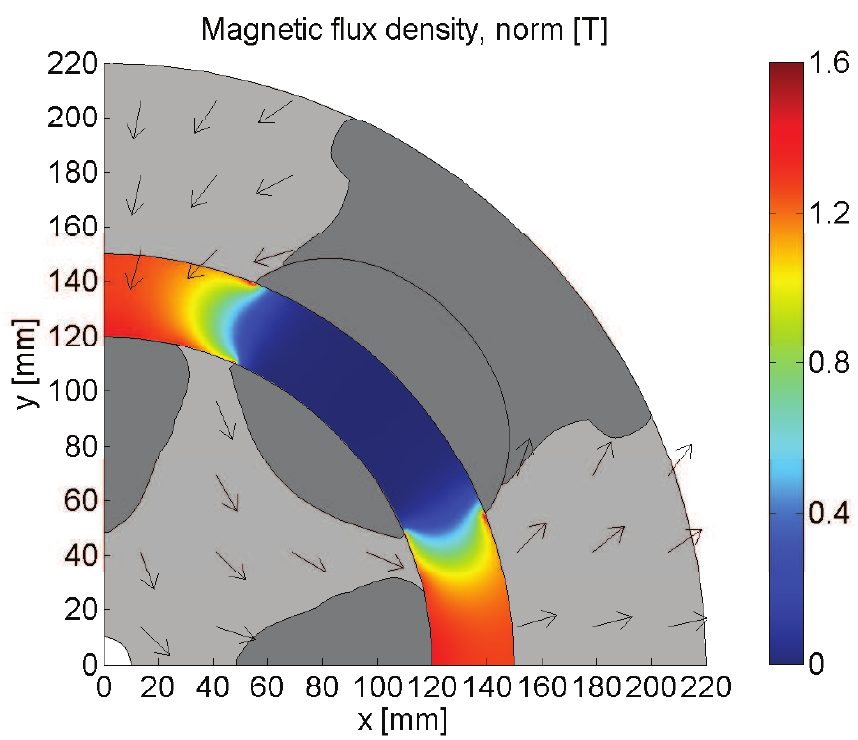}
\caption{Replacing magnet material with a high permeability soft magnetic material where the applied magnetic field is parallel or almost parallel to the remanence on the model shown in Fig. \ref{Fig.Optimization_scheme_pics_iron} (a) three successive times results in the magnet design shown. The line in the iron region in the outer magnet separates the iron regions generated by the improvement scheme and the  parallel replacement method and it is only shown for reference.}\label{Fig.Con_Hal_0-0-0-0-1-1-1-3_Imp}
\end{figure}

Having replaced magnet material by soft magnetic material according to Eq. (\ref{eq.HparallelBrNew}) and resolved the magnetic system, the magnet design must be investigated if there are now new regions where Eq. (\ref{eq.HparallelBrNew}) holds and magnet material can be replaced. The result of performing this replacement with a value of $\gamma = 0.075$ T to the model produced by a single application of the improvement scheme (Iron) is shown in Fig. \ref{Fig.Con_Hal_0-0-0-0-1-1-1-3_Imp}. This value of 0.075 T has been chosen such that magnet material is replaced in both the inner and outer magnet. For a lower value of $\gamma$ only material in the outer magnet is replaced. The optimal value of $\gamma$ would have to be analyzed for each individual magnet design. For this model the magnet material is replaced three successive times until the change in magnet volume from one iteration to the next is less than 5\%, at which point replacing the small remaining areas does not change the flux density significantly. The result of the replacement is also shown in Fig. \ref{Fig.Iterations_improvements}. By replacing magnet material with high permeability soft magnetic material the amount of magnet material is reduced by an additional 27\% compared to the original design while the difference in flux density was increased slightly by 4\%, again compared to the original design.

When replacing magnet material by soft magnetic material it is important to ensure that the shapes of the replaced segments are not such that the demagnetization of the segments are high as this can reduce the internal field in the soft magnetic material. However, as a high permeability material is used, even a very small field will cause the material to saturate, and thus this problem can be ignored except for cases with extremely high demagnetization.

\begin{figure}[!t]
\centering
\includegraphics[width=1\columnwidth]{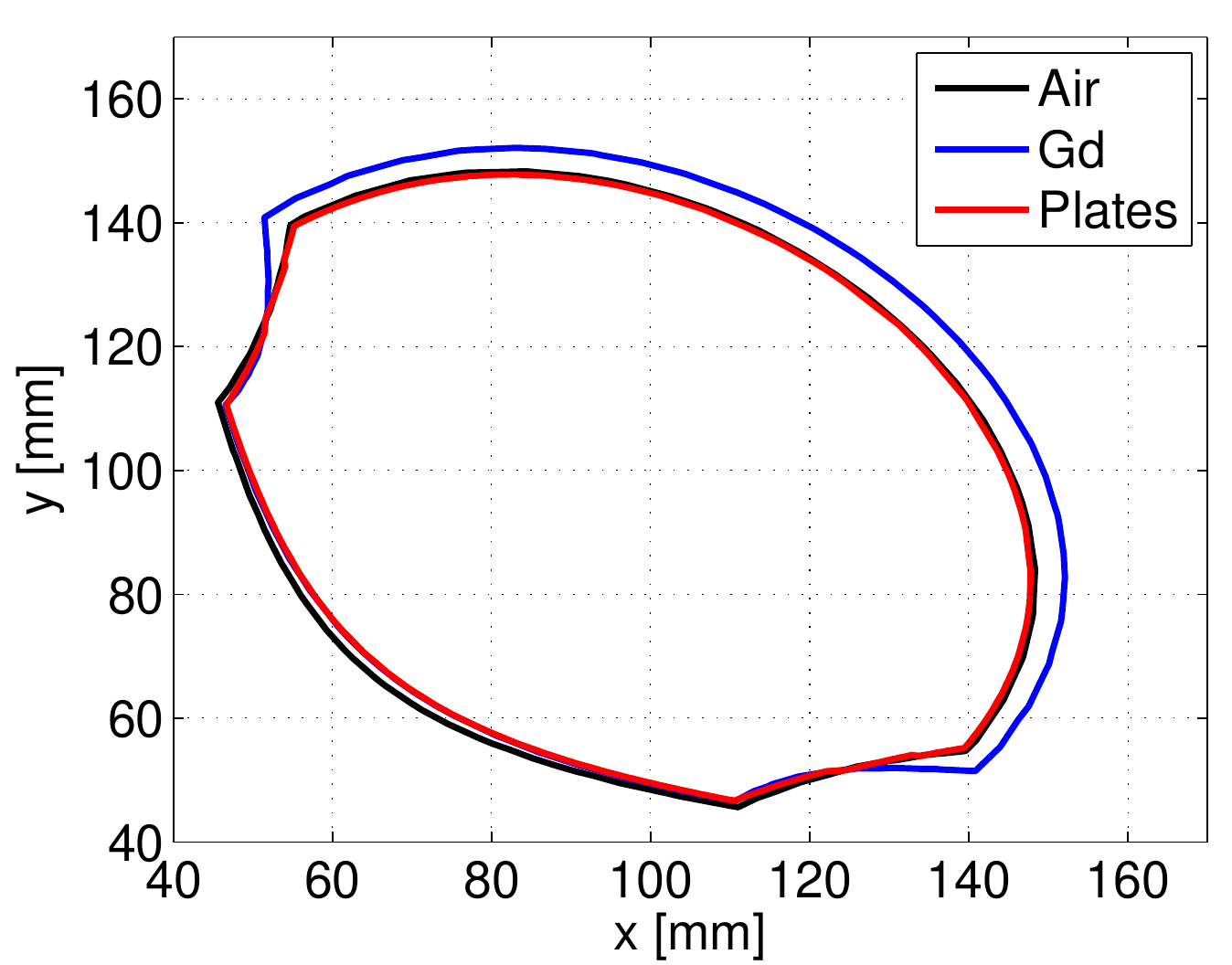}
\caption{The equipotential line of $A_\mathrm{z}$ that goes through the point $r = 135 \;\mathrm{mm}, \phi{} = 22.5^{\circ}$ for a system with an empty air gap, i.e. Fig. \ref{Fig.Con_Hal_0-0-0-0-1-0-0-1_Air_Air}b, and for a system where the air gap is completely filled with Gd or with 500 plates of Gd with a thickness of 0.9 mm. The permeability of Gd at 273 K has been used.}\label{Fig.Fig8}
\end{figure}

As can be seen from Figs. \ref{Fig.Iterations_improvements} and \ref{Fig.Con_Hal_0-0-0-0-1-1-1-3_Imp} replacing magnet material with soft magnetic material can also reduce the manufacturability of the magnet design, and thus the same consideration as with the improvement scheme applies. Here we will not consider segmenting the concentric Halbach cylinder design, as the design is only meant to serve as an example and also because no clear optimum segmentation procedure can be suggested. In Ref. \cite{Bjoerk_2010e} we apply the present improvement schemes to a magnet design which is then segmented and constructed. The resulting magnet show high performance for magnetic refrigeration.

When using a magnet design in an application a magnetic material will usually be placed within the air gap in the high and low field regions. This might alter the magnetic field in the high field region, which will alter the field lines and thus might lead to a different magnet design if the improvement scheme is applied with the material present in the air gap. However, unless high permeability materials are used, the magnetic field will change little. As an example consider magnetic refrigeration where Gd is commonly used as the magnetocaloric material that is placed in the air gap. This material has a relative permeability in the range of 2-10, depending on temperature and magnetic field. Using the permeability of Gd at 273 K as a function of magnetic field, as obtained from Ref. \cite{Bjoerk_2010d}, we have calculated the equipotential line of $A_\mathrm{z}$ for a case where the air gap was completely filled with Gd at a temperature of 273 K. For the first iteration this leads to a similar shaped equipotential line of $A_\mathrm{z}$, but which encloses an area that is 6.9\% larger than for the case without Gd. It is especially in the outer magnet that the equipotential line for the Gd case is larger than for the case with an empty air gap. For a packed sphere bed the porosity is usually around 0.36, which will lower the impact of placing Gd in the air gap. We have also calculated the equipotential line of $A_\mathrm{z}$ for a case where the air gap is filled with 500 plates of Gd. These plates have a thickness of 0.9 mm and a spacing in the center of the air gap of 0.8 mm. Using these plates the change in area of the equipotential line is only 2.9\% and the contours are almost identical. The equipotential line for both cases are shown in Fig. \ref{Fig.Fig8}. Thus placing a low permeability magnetic material in the air gap does not significantly change the equipotential line.

\section{Conclusion}
An algorithm for improving the difference in flux density between a high and a low flux density region in an air gap in a magnetic structure has been presented and as an example applied to a two dimensional concentric Halbach magnet design. For the design considered, applying the scheme reduces the amount of magnet material used by 15\% and increases the difference in flux density by 41\%. For the design considered here it was also shown that by replacing magnet material with a high permeability soft magnetic material where the applied magnetic field is parallel or almost parallel to the remanence the amount of magnet material used can be reduced by an additional 27\% compared to the original design while the difference in flux density was increased slightly by 4\%, again compared to the original design.

\section*{Acknowledgements}
The authors would like to acknowledge the support of the Programme Commission on Energy and Environment (EnMi) (Contract No. 2104-06-0032) which is part of the Danish Council for Strategic Research.

\end{document}